\documentclass[RNAAS]{aastex62}
\usepackage{amsmath}
\usepackage{natbib}

\begin{document}

\title{The Reionization Parameter Space Consistent with the Thomson Optical Depth from Planck}

\author[0000-0002-2411-2766]{Daniel Glazer}
\altaffiliation{dglazer@andrew.cmu.edu}
\affiliation{McWilliams Center for Cosmology, Department of Physics, Carnegie Mellon University\\ Pittsburgh, PA 15213, USA}

\author[0000-0003-3709-1324]{Markus Michael Rau}
\altaffiliation{markusmichael.rau@gmail.com}
\affiliation{McWilliams Center for Cosmology, Department of Physics, Carnegie Mellon University\\ Pittsburgh, PA 15213, USA}

\author[0000-0001-6778-3861]{Hy Trac}
\altaffiliation{hytrac@andrew.cmu.edu}
\affiliation{McWilliams Center for Cosmology, Department of Physics, Carnegie Mellon University\\ Pittsburgh, PA 15213, USA}

\keywords{}

\section{}
The Thomson optical depth $\tau$ quantifies the scattering of cosmic microwave background (CMB) photons by free electrons, and thus is an excellent probe of the Epoch of Reionization (EoR). Planck has recently presented an updated constraint of $\tau=0.054\pm0.007$ from measurements of the CMB temperature and polarization angular power spectra \citep{2018arXiv180706209P}. Previously, they inferred the midpoint of the EoR to be in the range $7.8 < z < 8.8$ for $\tau=0.058 \pm 0.012$, depending on the adopted model for reionization \citep{2016A&A...596A.108P}. However, the duration and the shape of the reionization history are poorly constrained using $\tau$ alone.

We study the EoR parameter space consistent with the current measurement of $\tau$ using a new parametrization of the reionization history by \citet{2018ApJ...858L..11T}. The evolution of the mass-weighted ionization fraction $x_\text{i}(z)$ is specified in terms of the redshift midpoint, duration, and asymmetry shape parameters. Let redshifts $z_{05}$, $z_{50}$, and $z_{100}$ correspond to ionization fractions $x_\text{i} = 0.05$, 0.50, and 1.00, respectively. We choose the midpoint as $z_{50}$ and define the duration and asymmetry as $\Delta_\text{z} = z_{05}-z_{100}$ and $A_\text{z} = (z_{05}-z_{50})/(z_{50}-z_{100})$, respectively. 
Differing from \citet{2018ApJ...858L..11T}, we use $z_{100}$ rather than $z_{95}$ to ensure that the ionization fraction reaches unity at the end of reionization across the entire parameter space of interest. Lagrange interpolating functions are then used to construct analytical curves for $x_\text{i}(z)$ that exactly fit the given ionization points.

We perform a simple MCMC analysis using conservative flat priors for the three shape parameters: $6 < z_{50} < 12$, $0 < \Delta_\text{z} < 20$, and $1 < A_\text{z} < 20$. A lower bound $z_{100} > 6$ is imposed, motivated by observations of the Lyman alpha forest toward the end of reionization \citep[e.g.][]{2015MNRAS.447..499M}. A lower bound $A_\text{z} > 1$ instead of 0 is motivated by the asymmetric reionization histories seen in radiation-hydrodynamic simulations \citep[e.g.][]{2017arXiv171204464D}. We calculate $\tau$ assuming standard cosmological parameters: $\Omega_\text{b}=0.045$, $\Omega_\text{m}=0.30$, $\Omega_\Lambda=0.70$, $h=0.7$, $Y=0.24$, and $z_\text{HeIII}=3$. We have checked the convergence of our chains by inspecting the traces and the Gelman-Rubin statistic. Note that a more rigorous analysis would start with the CMB temperature and polarization power spectra rather than the derived constraint on $\tau$.

\begin{figure}[t]
\begin{center}
\includegraphics[width=0.5\linewidth]{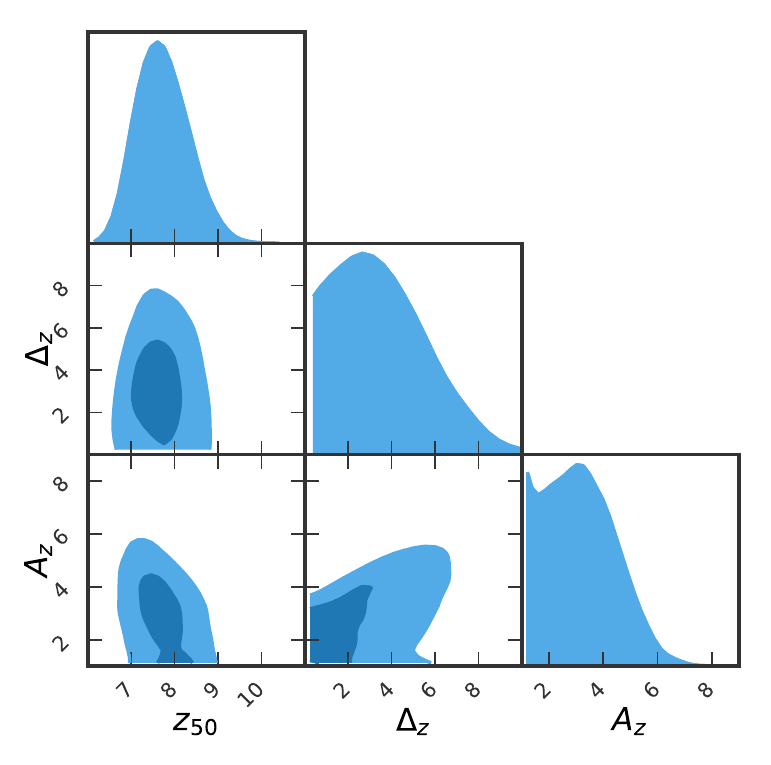}\includegraphics[width=0.5\linewidth]{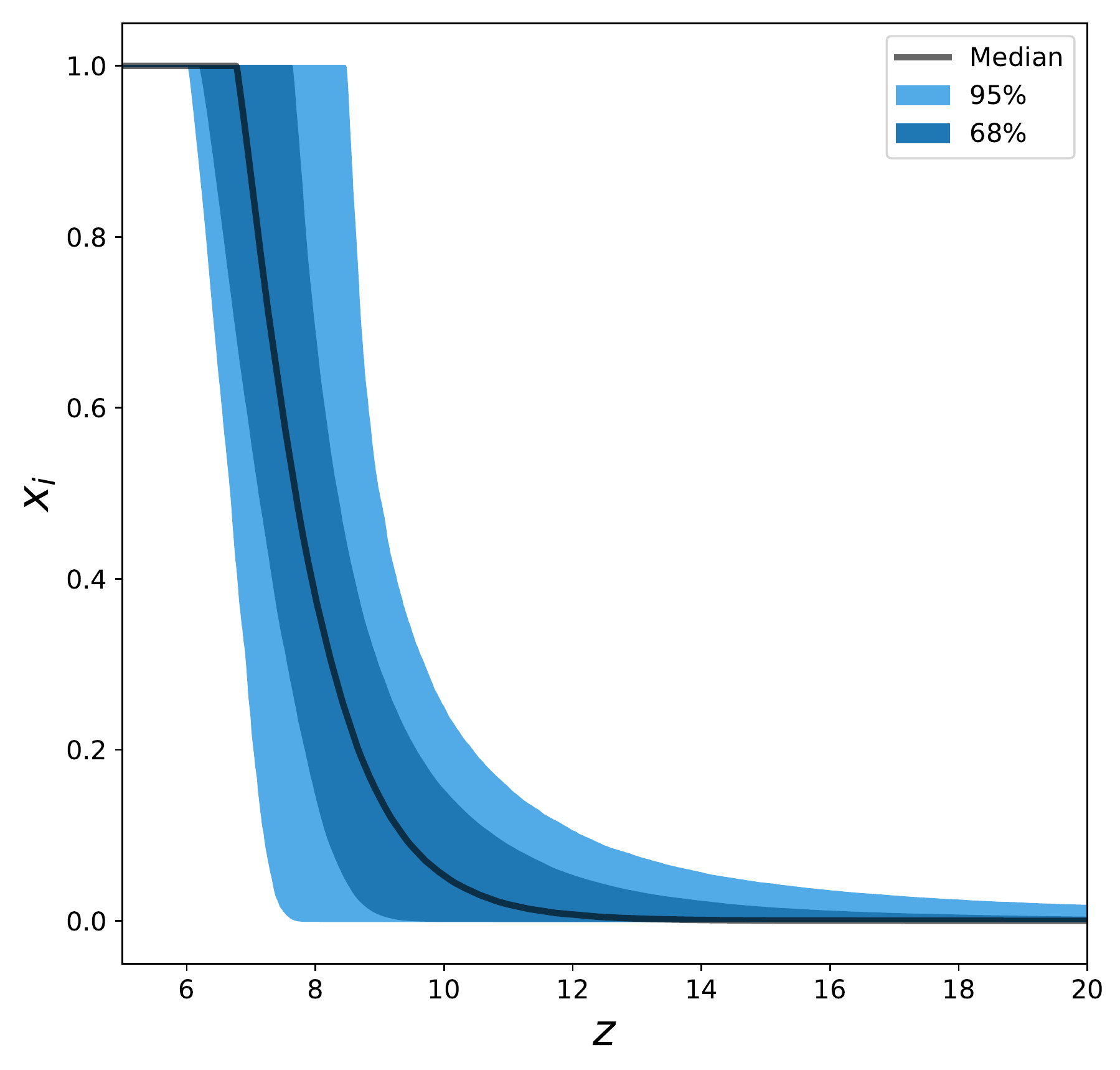}
\caption{\textit{Left:} Posterior distributions of the midpoint, duration, and asymmetry shape parameters. The top of each column shows the respective marginal distributions of the individual parameters. The contours show the 68\% (dark blue) and 95\% (light blue) parameter constraints. \textit{Right:} Point-wise confidence intervals for the evolution of the mass weighted ionization fraction with redshift. Shown are the median (black), 68th percentile (dark blue), and 95th percentile (light blue).
\label{fig:1}}
\end{center}
\end{figure}

Figure \ref{fig:1} shows the MCMC results for the three shape parameters and the ionization fraction. As expected, $\tau$ mainly probes the midpoint, which is constrained to be $z_{50}=7.66\substack{+0.57\\-0.61}\text{(68\%)}\substack{+1.26\\-1.02}\text{(95\%)}$. The duration and asymmetry are not well constrained, but we find 95\% upper limits of $\Delta_\text{z} < 7.5$ and $A_\text{z}<5.3$, respectively. The anticorrelation between $A_z$ and $z_{50}$ arises because a larger asymmetry corresponds to relatively earlier reionization, and therefore a later midpoint is necessary to have a fixed $\tau$. The positive correlation between $A_z$ and $\Delta_z$ arises because a more extended reionization has to be shifted towards higher redshifts in order to satisfy the $z_{100} > 6$ restriction for the end of the EoR. The small bimodal peak in the $A_z$ posterior distribution is also attributed to this restriction. Our reconstruction of the mass-weighted ionization fraction $x_\text{i}(z)$ appears similar to the FlexKnot \citep{2018arXiv180408476M} constraints by \citet{2018arXiv180706209P}.


\bibliography{Literature}
\bibliographystyle{aasjournal}

\end{document}